\documentclass[
aip,%
jmp,%
amsmath,%
amssymb,%
]{revtex4-1}

\usepackage{notoccite} % fixes numbering order in bibliography
\usepackage{amsmath}  % for math formulas
\usepackage{graphicx} % include figure files
\usepackage{float}    % forcing figure positions
\usepackage{csquotes} % match opening and closing quotation marks
\MakeOuterQuote{"}

\begin{document}

%\preprint{AIP/123-QED}

\title{Reconstruction in Time-Bandwidth Compression Systems}

\author{J. Chan}
\affiliation{ %
Department of Electrical Engineering, %
University of California, Los Angeles, CA 90095, USA
}
\author{A. Mahjoubfar}
\affiliation{ %
Department of Electrical Engineering, %
University of California, Los Angeles, CA 90095, USA
}
\affiliation{ %
California NanoSystems Institute, %
Los Angeles, California 90095 USA.
}
\author{M. Asghari}
\affiliation{ %
Department of Electrical Engineering, %
University of California, Los Angeles, CA 90095, USA
}
\author{B. Jalali} %
\email[Corresponding author: ]{jalali@ucla.edu}
\affiliation{ %
Department of Electrical Engineering, %
University of California, Los Angeles, CA 90095, USA
}
\affiliation{ %
Department of Bioengineering, %
University of California, Los Angeles, CA, 90095, USA
}

\date{\today}

\begin{abstract}
Recently it has been shown that the intensity time-bandwidth product of optical signals can be engineered to match that of the data acquisition instrument \cite{AST_AO,AST_APL,AST_Optica,BigData}. In particular, it is possible to slow down an ultrafast signal, resulting in compressed RF bandwidth - a similar benefit to that offered by the Time-Stretch Dispersive Fourier Transform (TS-DFT) - but with reduced temporal record length leading to time-bandwidth compression. The compression is implemented using a warped group delay dispersion leading to non-uniform time stretching of the signal's intensity envelope. Decoding requires optical phase retrieval and reconstruction of the input temporal profile, for the case where information of interest is resides in the complex field \cite{AST_AO,AST_APL,AST_Optica,BigData}. In this paper, we present results on the general behavior of the reconstruction process and its dependence on the signal-to-noise ratio. We also discuss the role of chirp in the input signal.
\end{abstract}

\maketitle

\section{Introduction}
\label{sec:intro}

Capturing wide-bandwidth ultrafast signals in real-time is restricted by the time-bandwidth product of the data acquisition instrument. Here, the "product" does not refer to the uncertainty limit but rather to that imposed by the digitizer's bandwidth and record length. Time Stretched Dispersive Fourier Transform (TS-DFT) has proven to be very useful in digitizing ultrafast optical signals in real-time. It has led to the discovery of Optical Rogue Waves \cite{Solli07}, the study of supercontinuum noise \cite{Wetzel12} and the detection of cancer cells in blood with one-in-a-million sensitivity \cite{Goda09,Goda12}. TS-DFT allows real-time analog-to-digital conversion by reducing the intensity (envelope) bandwidth of information-carrying optical signals to match the much smaller bandwidth of the electronic digitizer \cite{Goda13}. However, it does not address the generation of big data volumes inherent in high throughput real-time measurements. This has fueled interest in non-uniform time stretch transformations that take advantage of sparsity in physical signals to achieve both bandwidth compression as well as reduction in the temporal length \cite{AST_AO,AST_APL,AST_Optica,BigData}. Compared to linear time stretch achieved with TS-DFT, such transforms lead to savings in the amount of digital data required for transfer. The aim of this technique is to transform a signal such that its intensity matches not only the digitizer's bandwidth but also its temporal record length. The latter is typically limited by the digitizer's storage capacity.  To be sure, reduction of digital data size also facilitates its transmission over communication networks.

To this end, it was shown that it is possible to reshape the spectro-temporal profile of optical signals in the analog domain such that signal intensity's time-bandwidth product is commensurate with the acquisition instrument and the big data problem is alleviated \cite{AST_AO,AST_APL,AST_Optica,BigData}.  The compression is achieved through a type of time-stretch dispersive Fourier transform in which the transformation is intentionally warped using an engineered group delay (GD) dispersion profile. The transformation causes a frequency-dependent reshaping of the input waveform and, in the far field regime, can be interpreted as warped frequency-to-time mapping. The reconstruction method depends on whether the information of interest is in the spectral domain amplitude, or in the complex spectrum, i.e. whether the input temporal waveform needs to be recovered or not. If only the spectrum amplitude needs to be recovered, such as in the time stretch digitizer \cite{Bhushan98,Chou07} or camera \cite{Goda09,Goda12,ATOM,Qian09,Xing13}, reconstruction consists of a simple non-uniform time-to-frequency mapping using the inverse of the warped group delay function. Such a de-warping can be achieved using e.g. a look up table. When the information is in the temporal profile, reconstruction requires complex field recovery followed by digital backpropagation. To be compressed, the input signal must exhibit redundancy and sparsity \cite{AST_AO,AST_APL,AST_Optica,BigData}.

Previous reports have focused on reshaping of the signal intensity in frequency and time leading to time-bandwidth compression and expansion. In this report, we discuss the reconstruction process involving complex field recovery and backpropagation in the presence of noise. While the dispersive stretch transformation spans both near field and far field \cite{Goda13,Solli09,Asghari12}, in the case of near field ultrafast signals can be too fast to be digitized in real-time. The speed requirements of analog-to-digital converters may make the near field technique impractical for some applications.

\section{System Description}
\label{sec:system}

\begin{figure}[ht!]
  \centering
  \includegraphics[width=0.90\textwidth]{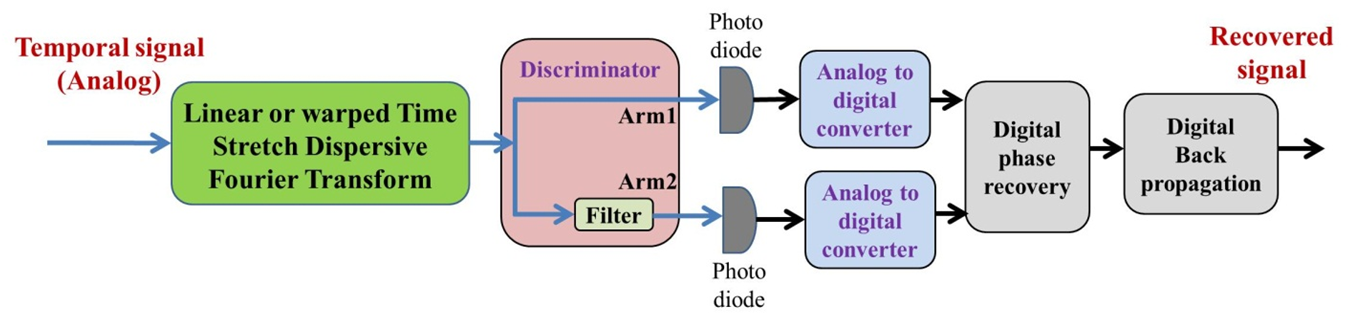}
  \caption{Block diagram of time bandwidth compression system using warped time stretch dispersive Fourier transform (TS-DFT). The analog waveform is first reshaped by warped time stretch transform. An analog discrimination filter is used for phase recovery. After photo detection and analog to digital conversion, the complex electric field is obtained using digital phase recovery algorithm and the input waveform is reconstructed using digital backpropagation.}
  \label{fig:block_diagram}
\end{figure}

Figure \ref{fig:block_diagram} shows the block diagram of the system investigated here. The encoding uses warped time-stretch dispersive Fourier transform with an engineered group delay, followed by intensity detection in the photodiode. The aim is to compress the intensity time bandwidth product to make digitization and storage of wideband optical waveforms possible in real-time and at high throughput. In the far field, the transform causes warped (nonlinear) frequency-to-time mapping. The decoding reconstructs the original waveform in the digital domain using complex field recovery followed by digital backpropagation through the warped group delay transform. The fidelity of the encoding process depends on the sparsity/redundancy of the input signal, while the decoding accuracy is influenced by the digital phase recovery and backpropagation algorithms and their tolerance to SNR. 

The amount of group delay dispersion used is dictated by the need to compress the main portion of the signal spectrum to the bandwidth of the ADC. The latter is the target bandwidth for time-bandwidth compression systems. In the case of time-bandwidth engineering, the warped shape of the group delay is further dictated by the desired output temporal duration. In other words, the kernel of the transform is designed based on the input signal bandwidth and duration, target output (digitizer) bandwidth and the desired output time duration. As previously demonstrated \cite{AST_AO,AST_APL,AST_Optica,BigData}, the information about the input signal bandwidth and spectral resolution as well as the desired output intensity bandwidth and ENOB of ADC are used to design the GD profile. Based on the sparsity of the signal under test, the amount of warping can be designed to achieve time bandwidth compression.

There are numerous techniques for recovering the complex amplitude from intensity-only measurements \cite{Solli09,Asghari12,Asghari12_2,Dorrer03,Walmsley09,Liu87,Grilli01,Zhang03,Chan08,Jaganathan12} and a review of the field is not the purpose of this paper. Here we use the STARS algorithm \cite{Asghari12} without having performed a comprehensive comparison with alternative techniques. While this technique may not be the optimum, the conclusions reached here are independent of the specific phase recovery techniques and apply, qualitatively, to other methods. The STARS technique obtains the complex-field using two intensity measurements \cite{Asghari12}. In particular, the signal is measured at the input and output of a linear optical discrimination filter, as shown in Figure \ref{fig:block_diagram}.

\section{SNR dependence of reconstruction}
\label{sec:snr_dep}
Reconstructing the temporal profile of the input waveform requires the full field of the dispersed signal and necessitates optical phase recovery. The ability to recover the optical phase from intensity measurements is influenced by the signal to noise ratio of the measurement. This issue is not unique to the present system and is true for all phase recovery techniques. The main sources of noise in high-speed measurements are thermal noise, shot noise, relative intensity noise and the quantization noise of the digitizer \cite{Mahjoubfar13}. The analog-to-digital converter (ADC) plays a central role in the system and its resolution and sampling rate (lower-bounded by the Nyquist rate) both affect the fidelity of the phase recovery and hence the reconstruction. The trade-off between ADC sampling rate and resolution places additional emphasis on the ADC. 

\begin{figure}[ht!]
  \centering
  \includegraphics[width=0.8\textwidth]{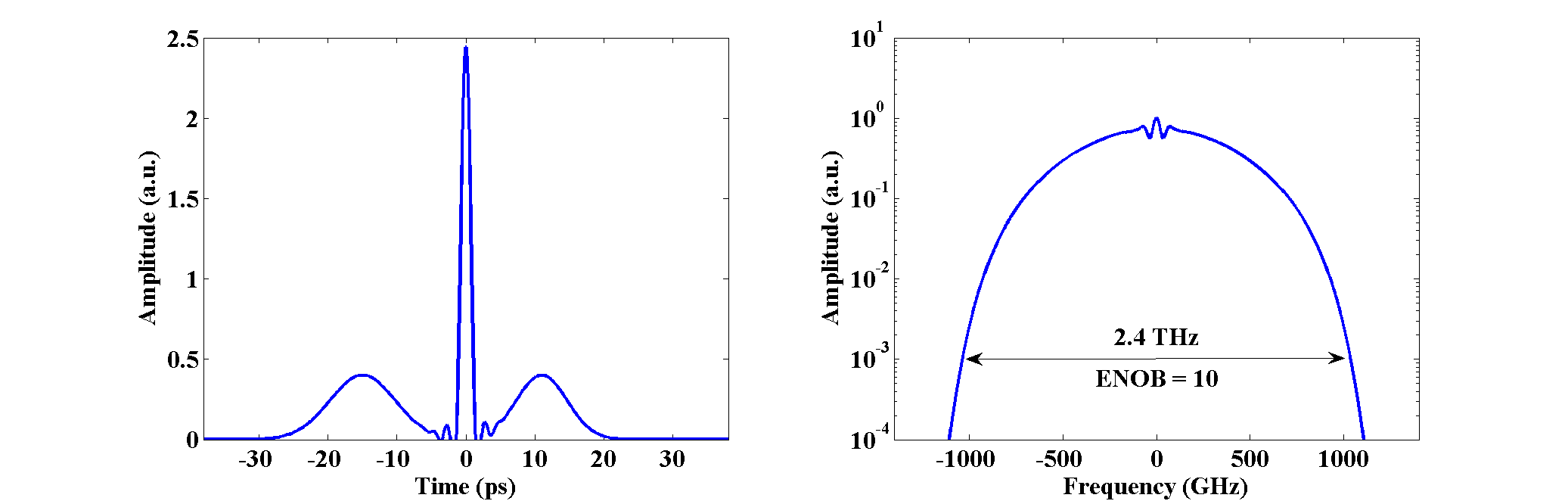}
  \includegraphics[width=0.8\textwidth]{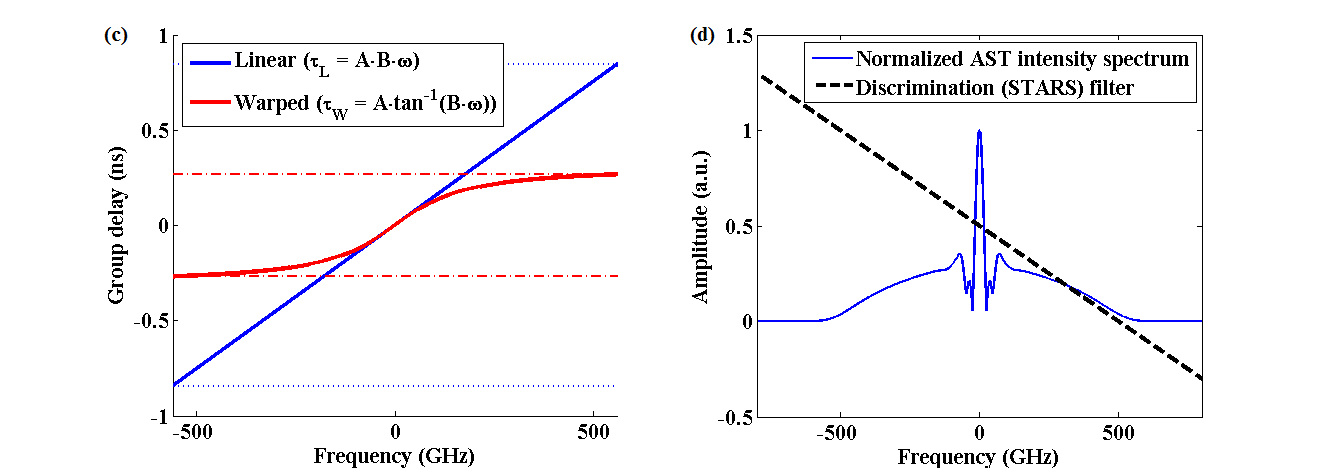}
  \caption{(a) Signal under test. (b) Intensity spectrum with 2.4 THz double sideband bandwidth (1.2 THz single sideband). (c) Group delay dispersion profiles for linear and warped stretches. The ratio of the dash-dot and dotted lines correspond to the designed time-bandwidth compression, and are referenced in figures \ref{fig:output}e and f. (d) Discrimination (STARS) filter amplitude, with the output intensity spectrum of the linear case as comparison.}
  \label{fig:input}
\end{figure}

To focus the discussion and give a specific example, we consider a test waveform shown in Fig. \ref{fig:input}a. The simulations include thermal noise and ADC quantization noise (measured as Effective Number of Bits, or ENOB); we assume a sufficiently high average pulse power such that shot noise is negligible relative to quantization noise. Figure \ref{fig:input}a and \ref{fig:input}b show the temporal profile and the spectrum of input signal under test. The input intensity single sideband bandwidth is roughly 1.2 THz (measured at -30 dB point).

We assume a digitizer bandwidth of 32 GHz and compare the linear stretch (TS-DFT) with the warped stretch (AST). Figure \ref{fig:input}c shows the group delay dispersion profiles used in each case and the legend shows their mathematical description. To achieve intensity time-bandwidth compression, the warped case uses a warped GD with sublinear frequency dependence. Figure \ref{fig:input}d shows the transfer function of the optical discrimination filter employed for phase recovery (Figure \ref{fig:block_diagram}) overlapped with the output signal spectrum.

\begin{figure}[H]
  \centering
  \includegraphics[width=0.78\textwidth]{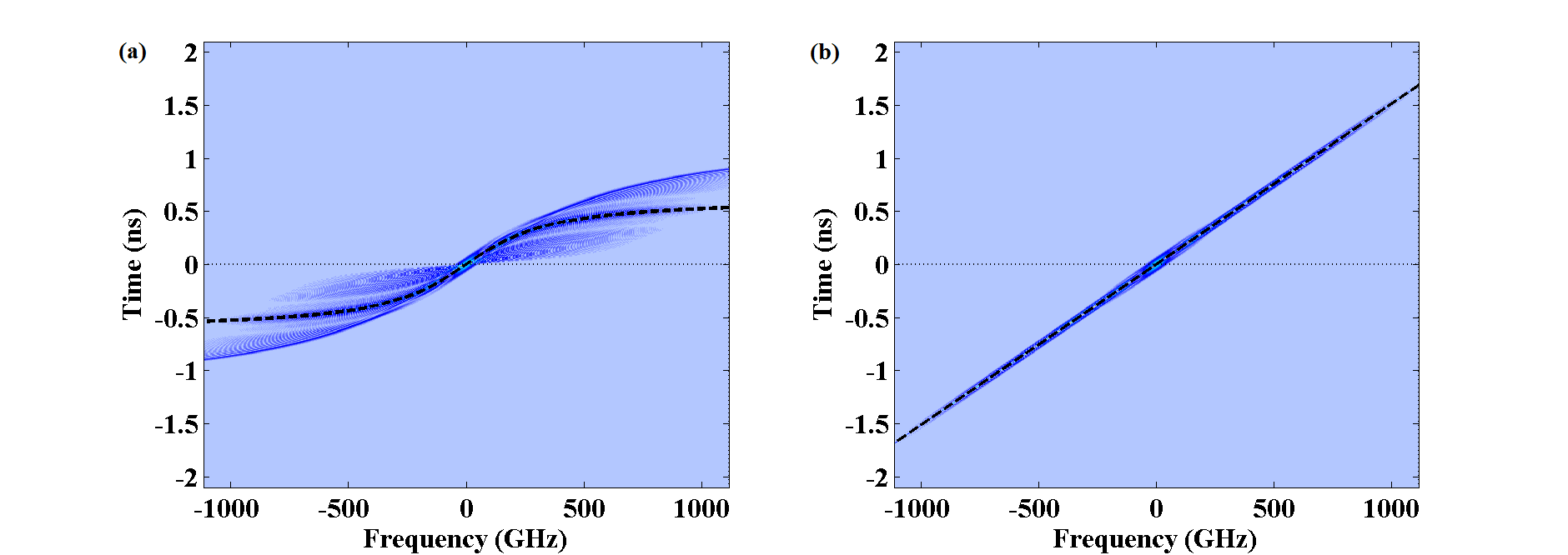}
  \includegraphics[width=0.78\textwidth]{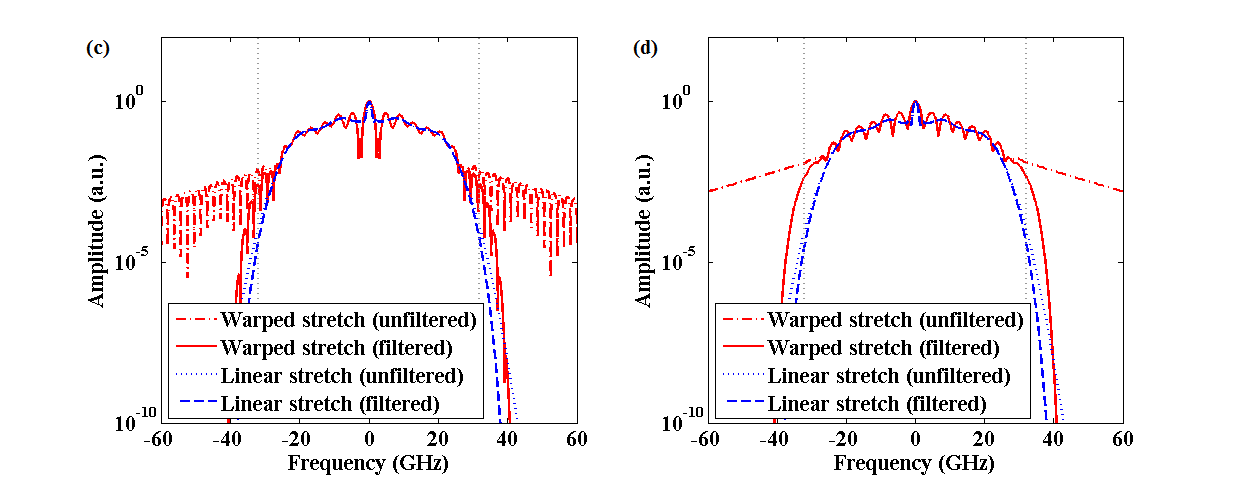}
  \includegraphics[width=0.78\textwidth]{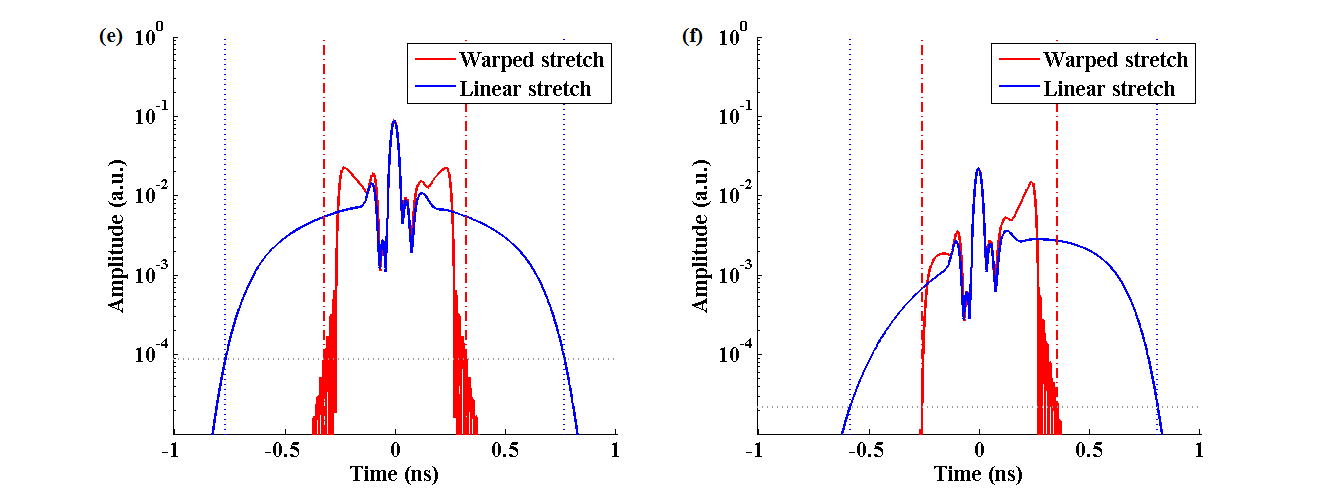}
  \caption{The Stretched Modulation Distribution ($S_M$) for the case of (a) warped stretch (b) linear stretch. (c) The RF spectrum of Arm 1 (see Fig. \ref{fig:block_diagram}) before and after application of the 32 GHz RF filter. (d) The same for Arm 2.  The corresponding output time domain signals after the photodiode (e) without and (f) with the discrimination filter. The 32 GHz RF filter is the digitizer bandwidth and the target bandwidth for the time-bandwidth compression system. The discrimination filter is used for phase recovery. The horizontal dotted line is the digitizer least significant bit (LSB) and defines the effective temporal duration, as shown by the vertical dotted lines. Note how figure \ref{fig:output}f differs from the ideal time duration as defined by the group delay in Figure \ref{fig:input}a, due to ENOB limitations.}
  \label{fig:output}
\end{figure}

The Stretched Modulation ($S_M$) function is a 3D distribution that visualizes the chirp caused by the linear and the warped GD. It plots the envelope intensity as a function of modulation frequency and the group delay. Figures \ref{fig:output}a and \ref{fig:output}b show the $S_M$ plots for the warped and linear transformed signals. The bandwidth of the central lobe in the spectrum is given by the intercept of the distribution with the horizontal axis and is determined by the slope of the GD dispersion near the origin. In both cases, the GD dispersion slope is designed to reduce the central lobe bandwidth to the target bandwidth of 32 GHz. The target bandwidth is ADC bandwidth and to model it a 32 GHz low pass filter is applied to the stretched signal before the quantization. This limits the signal bandwidth that is used for reconstruction to the designed (ADC) bandwidth.

Figure \ref{fig:output}c show the RF (intensity) spectra for both cases (linear and warped) before and after application of the 32 GHz anti-aliasing RF filter. We note that although the spectrum of the warped stretched waveform extends beyond the central lobe, the reconstruction only uses the portion of the spectrum within the compressed target (ADC) bandwidth. Also the same bandwidth is used for reconstruction in both the linear and the warped stretched signals. Figure \ref{fig:output}d shows the same for stretched output in Arm 2, i.e. after the discrimination filter (Figure \ref{fig:block_diagram}). The analog temporal signals, after the stretch transformation and application of the RF filter, are shown in figures \ref{fig:output}e (Arm 1) and \ref{fig:output}f (Arm 2). While the intensity bandwidth in both cases is compressed to 32 GHz, the temporal record length (sum of two arms) is reduced by 2.75 times for the warped stretch leading to the time-bandwidth compression by the same factor.

\begin{figure}[h!]
  \centering
  \includegraphics[width=0.5\textwidth]{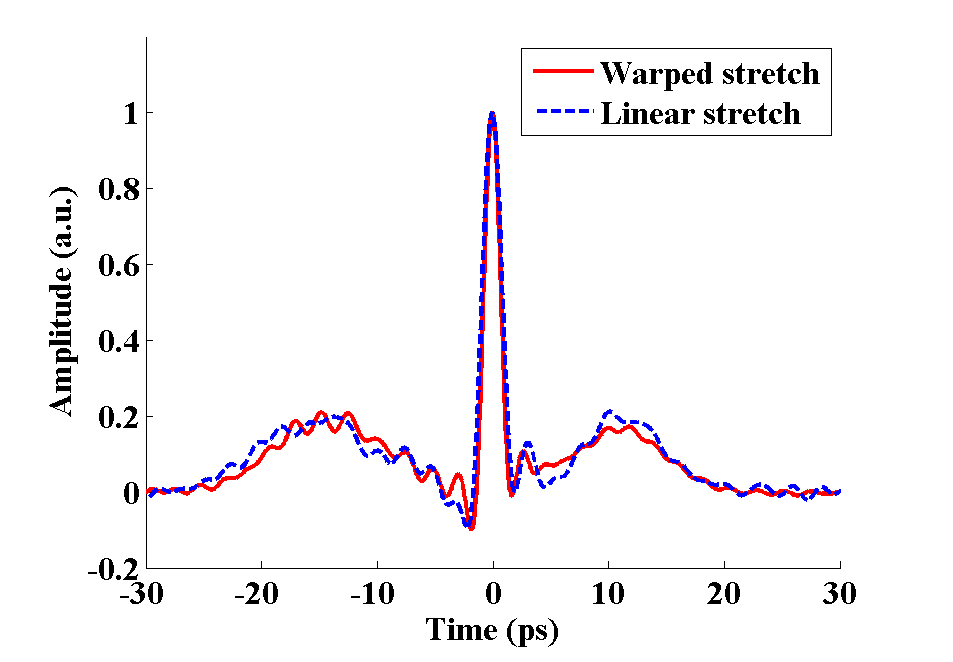}
  \caption{Reconstructed input signal after digital phase recovery and backpropagation for both linear and warped stretch cases. A digitizer with 10 ENOB was assumed.}
  \label{fig:reconstruction}
\end{figure}

The input temporal waveform is reconstructed through digital backpropagation using the measured amplitude and the recovered phase. Figure \ref{fig:reconstruction}a shows the reconstructed temporal waveform for the linear and warped stretched cases. The plots are for single shot capture and reconstruction. The accuracy of reconstruction depends on the SNR of the digitized signal. Playing a dominant role in the SNR, the ADC quantization noise depends on the resolution of the ADC and is reflected in the digitizer's Effective Number of Bits (ENOB). Figure \ref{fig:reconstruction} assumes 10 ENOB. The reconstruction accuracy is similar for both linear and warped cases. We note that without stretching (whether linear or warped), the input waveform would have been too fast to be digitized in real-time by the digitizer.

The applied GD chirps the input signal with a specific GD profile. From this point of view, the chirp in the input signal will also affect the compression, and will in turn dictate the design of the GD profile. To achieve time bandwidth compression, the GD profile should be the difference between the desired GD chirp and the chirp of the input signal.

\section{Summary}
\label{sec:summary}

Stretch transformations such TS-DFT (linear stretch) and AST (warped stretch) are useful for capture of wideband intensity modulated optical waveforms, such as for real-time instruments where high speed signals must be digitized in real-time. Compressing the intensity time-bandwidth product using warped group delay dispersion enables real-time digitization of an ultrafast signal and at the same time reduces the record length. Digital reconstruction of the original signal requires optical phase recovery in both the linear and warped stretch cases and its accuracy depends on the signal to noise ratio. For wideband systems, the SNR is typically limited by the resolution of the ADC \cite{Mahjoubfar13}. By way of example, we have simulated waveform reconstruction for the cases of warped stretch and linear stretch, where the warped case achieves a time-bandwidth compression of 2.75 times when compared to the linear case.

\section{Acknowledgement}
\label{sec:acknowledgement}

This work was supported by the Office of Naval Research (ONR) MURI Program, NECom.

\bibliographystyle{myIEEEtran}
\bibliography{20140901__ArXiv__AST_SINAD}

\end{document}